\documentclass[aip, jcp, amsfonts, amssymb, amsmath, reprint]{revtex4-1}

\usepackage{amsthm}
\usepackage[T1]{fontenc} 
\usepackage{amsfonts}
\usepackage{physics}
\usepackage{mathtools}
\usepackage{hyperref}  
\usepackage{xcolor} 
\usepackage{hyperref}  
\hypersetup{
    colorlinks = True,
    allcolors = {blue}
}
\usepackage{graphicx}
\usepackage{dcolumn}
\usepackage{breqn}
\makeatletter
\let\cat@comma@active\@empty
\makeatother

\usepackage{breqn}
\usepackage{natbib}
\usepackage{subfiles}

\def\LVac{\langle-|}
\def\RVac{|-\rangle}

\def\rAGP{|AGP\rangle}
\def\lAGP{\langle AGP |}
\def\C{c}
\def\G{{\Gamma}}
\def\P{{P}}
\def\Pdag{{P}^{\dagger}}
\def\N{{N}}
\def\prdm{\mathlarger{\gamma}}
\def\nrdm{\mathlarger{\nu}}
\def\K{{K}}
\def\z{ \mathlarger{Z} }
\def\sumESP{{\fontfamily{qcr}\selectfont sumESP}}
\def\poly{{\fontfamily{qcr}\selectfont poly}}

\usepackage{amsthm}
\usepackage[utf8]{inputenc}
\usepackage[english]{babel}

\usepackage[algoruled,boxed,lined]{algorithm2e}

\usepackage[caption=false]{subfig}

\begin{document}
\title{Efficient evaluation of AGP reduced density matrices }

\author{Armin Khamoshi}
    \email[Correspondence email address: ]{armin.khamoshi@rice.edu}
    \affiliation{Department of Physics and Astronomy, Rice University, Houston, TX 77005-1892}
    
\author{Thomas Henderson}
    \affiliation{Department of Chemistry, Rice University, Houston, TX 77005-1892}
    \affiliation{Department of Physics and Astronomy, Rice University, Houston, TX 77005-1892}    
    
\author{Gustavo Scuseria}
    \affiliation{Department of Chemistry, Rice University, Houston, TX 77005-1892}
    \affiliation{Department of Physics and Astronomy, Rice University, Houston, TX 77005-1892}

\date{\today} 

\begin{abstract}
We propose and implement an algorithm to calculate the norm and reduced density matrices of the antisymmetrized geminal power (AGP) of any rank with polynomial cost. Our method scales quadratically per element of the reduced density matrices. Numerical tests indicate that our method is very fast and capable of treating systems with a few thousand orbitals and hundreds of electrons reliably in double-precision. In addition, we present reconstruction formulae that allows one to decompose higher order reduced density matrices in terms of linear combinations of lower order ones and geminal coefficients, thereby reducing the computational cost significantly.
\end{abstract}


\maketitle

\section{Introduction} \label{sec:intro}

In electronic structure theory, a geminal is a wavefunction for two electrons. Geminals are central to the concept of bonding and have a long
history in quantum chemistry. \cite{surjan_introduction_1999} The geminal creation operator can, in general, be written as
\begin{equation}
\G^{\dagger }=\sum_{pq}^{2M}\eta _{pq} \C_{p}^{\dagger } \C_{q}^{\dagger },
\end{equation}%
where $\eta _{pq}$ is antisymmetric, $2M$ is the total number of spin-orbitals and $ \C_{p}^{\dagger }$ is the creation operator of a fermion in spin-orbital $p$.  

The simplest geminal wave function is perhaps the antisymmetrized geminal power (AGP) where all geminals are the same \cite{coleman_structure_1965}
\begin{equation}
\left\vert AGP\right\rangle =\frac{1}{N!}\left( \G^{\dagger }\right)
^{N}\left\vert -\right\rangle.
\end{equation}%
Here $\left\vert -\right\rangle $ is the physical vacuum containing no
electrons, $N$ is the number of pairs (2$N$ electrons), and the factor $1/{N!}$ is introduced for convenience. 

Without loss of generality, we choose to work in the natural orbital basis of the geminal. This is accomplished by applying a unitary transformation that brings $\eta_{pq}$ into a block diagonal form, wherein the one-body density matrix is diagonal and all spin-orbitals are paired. In this basis, it is mathematically simpler to work with hardcore boson operators, so that
\begin{equation}
\G^{\dagger }=\sum_{p=1}^{M}\eta _{p}\Pdag_{p},
\end{equation}%
where
\begin{eqnarray}
\Pdag_{p} &=& \C_{p}^{\dagger }\C_{\bar{p}}^{\dagger } \\
\N_{p} &=& \C_{p}^{\dagger } \C_{p}+ \C_{\bar{p}}^{\dagger } \C_{\bar{p}}. 
\end{eqnarray}
such that $ \C_{p}^{\dagger }$ is the fermion creation operator in orbital $p$, and $\bar{p}$ is the "paired" companion of $p$. The pair creation and annihilation operators, $\Pdag_{p}$, and $\P_{p}$ along with $\N_p$ are the generators of a global $su(2)$ algebra
\begin{eqnarray}\label{eq:CommutationRelations}
\left[ \P_{p}, \Pdag_{q}\right]  &=&\delta _{pq}\left( 1- \N_{p}\right) 
\\
\left[ \N_{p}, \P_{q}^{\dagger }\right]  &=&2\delta _{pq} \Pdag_{q}.
\nonumber
\end{eqnarray}
Notice that, written in its natural orbital basis, AGP exhibits the so-called \textit{seniority} symmetry. That is, orbitals with bars and no bars are either both occupied or empty. While in a general Hamiltonian the seniority-defining pairing scheme is arbitrary but can be optimized, in Hamiltonians for which seniority is a symmetry the pairs are naturally defined.

In the nuclear structure and condensed matter physics communities, AGP is known as the number projected Bardeen-Cooper-Schrieffer (PBCS) wavefunction. \cite{ring_nuclear_1980, blaizot_quantum_1986} The claim to fame of AGP is its ability to describe off-diagonal long range
order, a criterion for superconductivity, \cite{yang_concept_1962} without breaking
number symmetry as in BCS. \cite{bardeen_theory_1957} AGP is not a great wavefunction per se in quantum chemistry or condensed matter physics, because in most situations electron pairs are very different from each other. However, AGP
is potentially an excellent starting point for geminal correlation models
because it encompasses a combinatorial number of Slater determinants.
Indeed, it is easy to see that
\begin{equation}
    \rAGP = \underbrace{ 
    \sum_{ 1 \leq p_1<...<p_N \leq M} \eta _{p_1}...\eta _{p_N}\Pdag_{p_1}...\Pdag_{p_{N}}}_{ \textit{${M \choose N}$ terms}} \RVac.
\end{equation}%
This is a superposition of all possible seniority zero (paired) determinants involving $N$ pairs in $M$ orbitals but the coefficients are factorized by $\eta_{p_i}$ rather than being a tensor. In the latter case, when coefficients are general rather than factorized, the wavefunction is known as doubly occupied
configuration interaction (DOCI). \cite{veillard_complete_1967, couty_generalized_1997, kollmar_new_2003, bytautas_seniority_2011} DOCI is obtainable by exact diagonalization over
the space of seniority zero determinants with combinatorial cost as
a function of $M$ and $N$. However, AGP (PBCS) can be solved
variationally and optimized using symmetry breaking and restoration
techniques with mean-field $\mathcal{O}\left( M^{3}\right)$ cost. \cite{sheikh_symmetry-projected_2000, scuseria_projected_2011} Other classes of geminal theories like Bethe ansatz (BA) for solving the integrable Richardson-Gaudin Hamiltonians \cite{richardson_restricted_1963, dukelsky_colloquium:_2004, johnson_size-consistent_2013} or APIG (antisymmetrized product of interacting geminals) wherein all geminals are different \cite{limacher_new_2013} are very interesting models but have combinatorial
cost when applied to general (non-integrable) Hamiltonians. 

AGP is qualitatively correct for attractive pairing interactions at all correlation
regimes where many other models have serious difficulties. \cite{henderson_quasiparticle_2014-1, henderson_pair_2015, degroote_polynomial_2016, qiu_particle-number_2019} This makes AGP an attractive starting point for more accurate
correlated geminal theories. Recent work in our research group \cite{henderson_geminal-based_2019} has shown that an AGP-based configuration interaction (CI) is a promising step in this direction. For AGP-based correlated theories to be computationally affordable for large systems, it is crucial that the AGP reduced density
matrices (RDMs), loosely defined here as the expectation value of strings of
ordered generators $\left\langle \Pdag_{p}...\N_q...\P_{r}...\right\rangle $ over AGP, be obtainable with low cost. This is a necessary and crucial ingredient for developing successful geminal theories based on AGP.

In order to meet this goal, we introduce two techniques in this paper. First, we develop an efficient way of calculating the individual elements of RDMs to all ranks. For this, we formulate all RDMs in terms of the elementary symmetric polynomials and use the \sumESP{} algorithm \cite{gerhard_h._fischer_einfuhrung_1974, rehman_computing_2011, jiang_accurate_2016} to compute the sum efficiently. We argue that, with appropriate normalization, this is a reliable, fast, and stable method capable of treating large systems. Secondly, we rigorously prove that all AGP RDMs are expressible in terms of linear combinations of lower rank RDMs and geminal coefficients using what we call \textit{reconstruction formulae}. These formulae are exact and do not rely on cumulant decomposition of density matrices. As such, this is the most significant and novel contribution of this paper and has important theoretical and numerical implications. The significance relies on the fact that all correlated theories require high rank RDMs---often as high as 5- and 6-body. Therefore the ability to break down high-rank RDMs makes AGP a good starting point for correlated methods. Indeed, our result is reminiscent of Hartree-Fock theory wherein all high rank RDMs can be written as products of 1-body RDMs. From a computational perspective, the reconstruction formulae presented here reduce the cost and scaling of correlated AGP calculations significantly as demonstrated in Sec. \ref{sec:Energy}.

\section{Basic expressions} \label{sec:Background}

The analytic expressions for the norm and RDMs of AGP already exist in the literature and have been expressed in many different forms. \cite{dietrich_conservation_1964, coleman_structure_1965, ma_microscopic_1977, weiner_calculation_1980, ortiz_agp_1981, cioslowski_many-electron_2000} For the sake of completeness, and to familiarize the reader with our notation, we present our own version of these derivations. In so doing, we introduce a Lie algebraic approach to understanding how $\Pdag_p$, $\P_p$, and $\N_p$ act on the manifold of AGP and its excitations. And, we purposefully formulate all the matrix elements in such a way that they can be directly computed by the elementary symmetric polynomials. 

\subsection{Norm of the AGP wavefunction}
The norm of the AGP wavefunction corresponding to $N$ pairs and $M$ orbitals can be obtained by calculating the contractions explicitly. From the commutation relations in Eq. (\ref{eq:CommutationRelations}) and the fact that $(\P_{q_i})^2 = 0$, it follows that
\begin{align}
\begin{split}
    \lAGP AGP \rangle = \nonumber \\ 
\end{split} \\    
\begin{split}
    \sum_{\substack{1 \leq p_1<...<p_N \leq M \\ 1 \leq q_1<...<q_N \leq M}} \eta_{p_1}...\eta_{q_N} \LVac \P_{p_1}...\P_{p_N} \Pdag_{q_1}...\Pdag_{q_N} \RVac \nonumber
\end{split}\\
\begin{split}\label{eq:AGPoverlap}    
    = \sum_{1 \leq p_1<...<p_N \leq M} \eta_{p_1}^2...\eta_{p_N}^2 \equiv S_N^{M},
\end{split}
\end{align}
where $S_N^{M}$ is the elementary symmetric polynomial (ESP) of degree $N$ with $M$ variables associated with the vector $\{ \eta_1^2,..., \eta_M^2 \}$. 

It is often convenient and numerically better posed to work with normalized AGP, i.e.  $\lAGP AGP \rangle = 1$. One can easily verify that the following choice does the job:
\begin{equation} \label{eq:normalizedETA}
    \eta_p \xrightarrow{}  \frac{\eta_p}{ (S^{M}_N)^{ \frac{1}{2N} } }.
\end{equation}

\subsection{Differential Representation} \label{sub:Diff_rep}
 Before we derive the expressions for the RDMs, we need to understand how $\Pdag_p$, $\P_p$, and $\N_p$ act on AGP. The AGP state and its excitations describe a $Riemann$ manifold with a positive semidefinite metric. In Lie algebra terms \cite{gilmore_lie_2008} this implies that generators acting on AGP may be represented as differential operators. By direct calculation, one can show that
 \begin{equation}
     \left[ \N_{p},\left( \G^{\dagger }\right) ^{N}\right] = 2N \eta_p \left( \G ^{\dagger }\right)^{N-1}\Pdag_{p}.
 \end{equation}
 from which we get%
 \begin{equation} 
     \N_{p}\left\vert {N} \right\rangle = 2\eta _{p} \Pdag_{p}\left\vert {N}-1\right\rangle,
    \end{equation}
where $| {N} \rangle$ is the shorthand notation for $\rAGP$ with $N$ pairs. On the other hand, using explicit derivatives with respect to $\eta _{p}$ on 
$\left( \G^{\dagger }\right) ^{N}$, one obtains 
\begin{equation} \label{eq: Pdag_diff}
    \frac{\partial }{\partial \eta _{p}}\left\vert {N} \right\rangle =\Pdag_{p}\left\vert {N}-1\right\rangle.    
\end{equation}
This results in%
\begin{equation} \label{eq: N_diff}
    \N_{p}\left\vert N \right\rangle = 2\eta _{p}\frac{\partial }{\partial \eta_{p}}\left\vert N\right\rangle .   
    \end{equation}
A similar derivation for $\P_{p}$ yields%
\begin{eqnarray}
    \left[ \P_{p},\left( \G ^{\dagger }\right) ^{N} \right] = N \eta_p \left(
    \G ^{\dagger }\right) ^{N-1} \left( 1-\N_{p}\right) \nonumber \\
    - N \left( N-1 \right) \eta _{p}^{2} \Pdag_p \left( \G
    ^{\dagger }\right) ^{N-2},
\end{eqnarray} 
and therefore%
\begin{eqnarray}
    \P_{p}\left\vert N \right\rangle &=& \eta _{p}\left\vert
    N-1\right\rangle -\eta _{p}^{2} \Pdag_p \left\vert N-2\right\rangle \nonumber \\
     &=& (\eta _{p} - \eta _{p}^{2}\frac{%
    \partial }{\partial \eta _{p}}) \left\vert N-1\right\rangle.  \label{eq:P_diff}
\end{eqnarray}
Eq. (\ref{eq: N_diff}), and Eq. (\ref{eq:P_diff}) are the differential representation of operators $\N_p$ and $\P_p$ over AGP. Noting that $\Pdag_p$ is the Hermitian conjugate of $\P_p$ and acts to the left, we have the differential representation for all the generators as desired. As a corollary to these equations and the commutations relations of Eq. (\ref{eq:CommutationRelations}), it is easy to show that $(\N_p)^2 = 2 \N_p$ and $\N_p = 2\Pdag_p \P_p$ over AGP. We frequently use these properties in this paper. 

\subsection{AGP reduced density matrices} \label{sub:epRDM}
Consider a many-body system of fermions. To evaluate the energy or other observables thereof, one needs to compute the many-body RDMs. Since AGP exhibits seniority symmetry and the total number of pairs is fixed, only two kinds of contractions are nonzero
\begin{subequations}
\begin{gather}
    \begin{split} \label{eq:contraction1}
    \lAGP... \C^{\dagger}_{p} \C_{p}...\rAGP 
    \end{split}\\
    \begin{split} \label{eq:contraction2}
        \lAGP... \C^{\dagger}_{p} \C^{\dagger}_{\bar{p}} \C_{q} \C_{\bar{q}}...\rAGP.   
    \end{split}
\end{gather}
\end{subequations}
Clearly, all terms like that of Eq. (\ref{eq:contraction2}) can be written as $\Pdag_p$ and $\P_q$ by definition; and all like that of Eq. (\ref{eq:contraction1}) can be replaced by $\N_p/2$ as all electrons come in pairs. Therefore, all electron density matrices can be written as linear combinations of terms like $\left\langle \P_{p}^{\dagger}...\N_q...\P_{r}...\right\rangle $. Recall that $\N_p = 2 \Pdag_p \P_p$ over AGP; therefore, we can ultimately normal order everything such that all $\Pdag$ are on the left and all $\P$ are on the right, i.e. $\langle \Pdag_p...\P_q... \rangle$. We refer to these density matrices as \textit{pair} RDMs. 

    We set out by getting the expression for the $n=1$ pair RDM
    \begin{equation}
        \prdm^p_q = \lAGP \Pdag_p \P_q  \rAGP.
    \end{equation}
    Notice that    
    \begin{align}
        \begin{split}
            \P_q \rAGP = (\eta _{q} - \eta _{q}^{2}\frac{%
            \partial }{\partial \eta _{q}}) \left\vert N-1\right\rangle 
        \end{split} \nonumber\\
        \begin{split}
            = \eta_q \sum_{ \substack{p_1 <...<p_{N-1} \\ p_i \neq q} }^{M-1} \eta_{p_1}...\eta_{p_{N-1}} \Pdag_{p_1}...\Pdag_{p_{N-1}}  \RVac,
        \end{split} 
    \end{align}
    and from this it follows
    \begin{eqnarray}
        \prdm^p_q &=&  \eta_p \eta_q \sum_{ \substack{p_1<...<p_{N-1} \\ p_i \neq p,q} }^{M-2} \eta_{p_1}^2...\eta_{p_{N-1}}^2 \nonumber \\
        &=& \eta_p \eta_q S_{N-1}^{M-2} (\eta_p^2, \eta_q^2),
    \end{eqnarray}
    where $S_{N-1}^{M-2} (\eta_p^2, \eta_q^2)$ denotes ESP over $\{ \eta_1^2,..., \eta_{M-2}^2 \}$ such that $\eta_p^2$ and $\eta_q^2$ are omitted. Obviously if $p=q$, then $\prdm^p_p = \eta_p^2 S_{N-1}^{M-1}(\eta_p^2).$\\    
    
    Generalization of this to higher rank RDMs follows the same reasoning. By induction on $n$, one can show that the matrix elements of the $n$-pair RDM, $\mathlarger{\gamma}^{(n)}$, is
    \begin{eqnarray}\label{eq:npRDM}
         \prdm^{p_1<...<p_n}_{q_1<...<q_n} &=&  \lAGP \Pdag_{p_1}...\Pdag_{p_n} \P_{q_1}...\P_{q_n}  \rAGP \nonumber\\
        &=& (\prod_{i=1}^{n} \eta_{p_i}\eta_{q_i})S_{N-n}^{M-d} (\eta^2_{p_1},...,\eta^2_{q_n}),
    \end{eqnarray}
    where we used the facts that $\prdm^{p_1...p_n}_{q_1...q_n}$ is symmetric with respect to permutations of $p_i$'s and $q_i$'s, and that if $p_i = p_j$ or $q_i = q_j$ for some $i,j$ then the corresponding matrix element is instead zero. Here, $d = 2n - |\{ p_i \} \cap \{ q_i \}|$ which counts the number of unique indices among $p_i$'s and $q_i$'s. There is an extra symmetry in Eq. (\ref{eq:npRDM}); $p_i$ is interchangeable with $q_j$ for some $i,j$ if and only if $p_i \notin \{q_1...q_{j-1}q_{j+1}...q_n\}$ and $q_j \notin \{p_1...p_{i-1}p_{i+1}...p_n\}$. We make use of this property later in Sec. \ref{sec:reconstruction}.    
    
    There is a special case of ${\prdm}^{p_1...p_n}_{q_1...q_n}$  that we need later in this paper; that is when $p_i = q_i$ for all $i$. Again, by $\N_p = 2 \Pdag_p \P_p$, this can be written in terms of $\langle \N_{p_1} \N_{p_2} ... \N_{p_n} \rangle$. We refer to this as a \textit{number} RDM. Formally, we define the number RDM of rank $n$, $\nrdm^{(n)}$, as
    \begin{eqnarray}
        \nrdm_{p_1...p_n}= \lAGP \N_{p_1} \N_{p_2}... \N_{p_n} \rAGP,
    \end{eqnarray}    
    where all the indices are assumed to be different---otherwise it would be equal to a lower rank number RDM by $\N_p^2 = 2\N_p$. From Eq. (\ref{eq:npRDM}) this can be computed by
    \begin{equation}\label{eq:nNRDM}
         \nrdm_{p_1...p_n} =(\prod_{i=1}^{n} 2\eta_{p_i}^2)S_{N-n}^{M-n} (\eta^2_{p_1},...,\eta^2_{p_n}).
    \end{equation}    

\section{Numerical Algorithm} \label{sec:Algorithm}

\begin{figure}[t]
    \centering
    \includegraphics[width=0.45\textwidth, keepaspectratio]{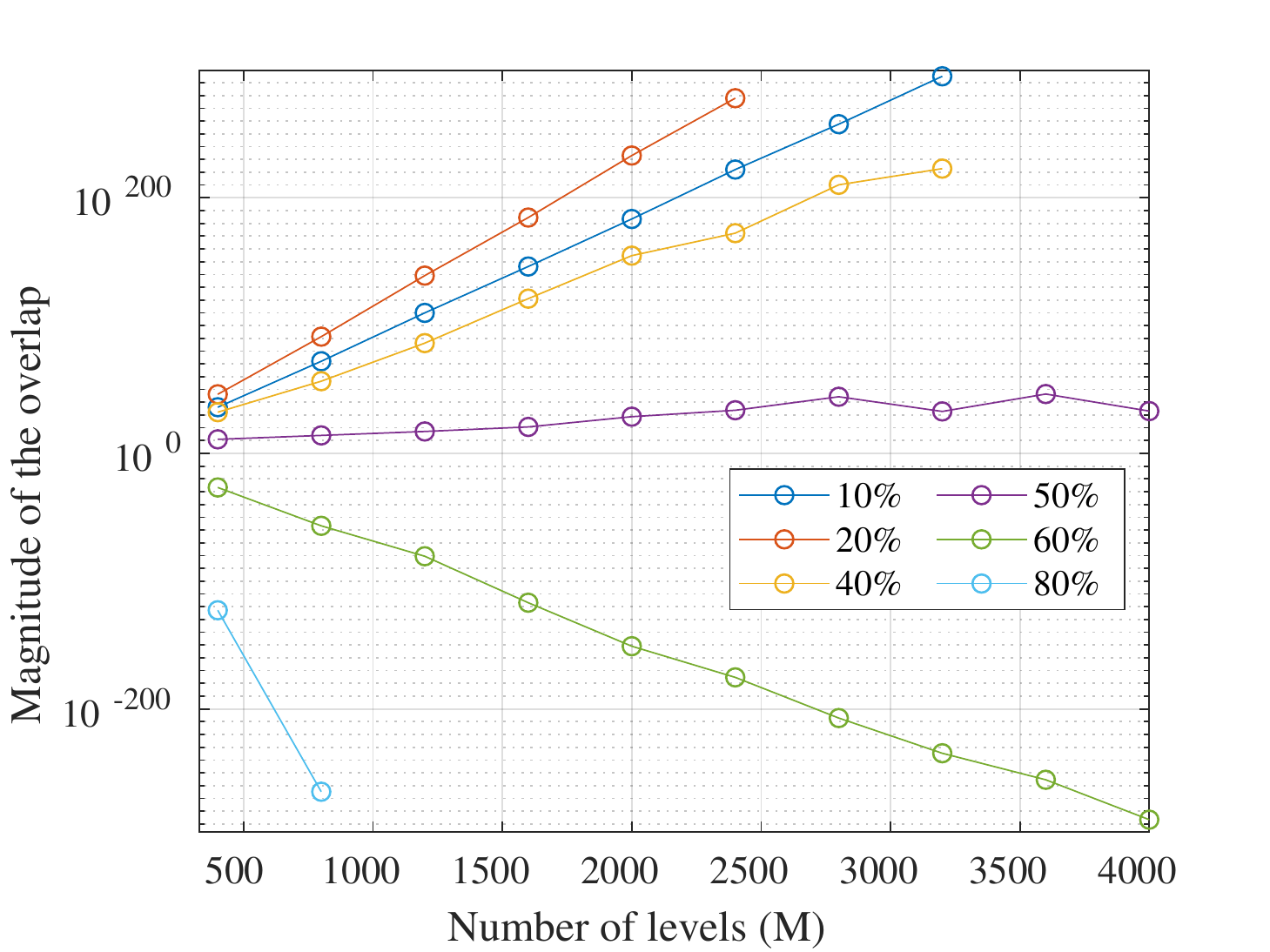}
    \caption{Magnitudes of the norm of AGP as a function of number of orbitals for randomly selected geminal coefficients. Each plot is for different values of $N$ reported as the percentage of $M$. Here, each point is the sample mean of 10 observations taken from $\eta_p \sim \text{Unif}(0,1)$.}
    \label{fig:overflow_underflow}
\end{figure}       

From Sec. \ref{sec:Background} it is clear that computing ESP efficiently is imperative to using AGP for any realistic system. This is because there are ${M \choose N}$ summands in every $S_N^{M}$ and a straightforward summation could grow combinatorially with system size. Elegant analytical formulation of ESP such as the one introduced in Ref. \cite{lee_power_2016} also scale combinatorially. In this section, we introduce the \sumESP{} algorithm \cite{jiang_accurate_2016} that calculates ESP with polynomial cost. First, we briefly review some important remarks about the error analysis that were extensively studied in Ref. \cite{rehman_computing_2011, jiang_accurate_2016} In Sec. \ref{sub:IdivnRDM_cost} and \ref{sub:FullnRDM_cost}, we study the time scales of various computations involving the norm of AGP and the corresponding RDMs as a function of $N$ and $M$. The computer environments used for the runtime measurements can be found in Appendix \ref{sec:Computer_env}.

\subsection{\sumESP{} and error analysis}\label{sub:NumAnlysis}
    \begin{figure*}[t]
        \centering
        \subfloat[\label{fig:1000L_overlap}]
            \centering
            {\includegraphics[width=0.45\textwidth]{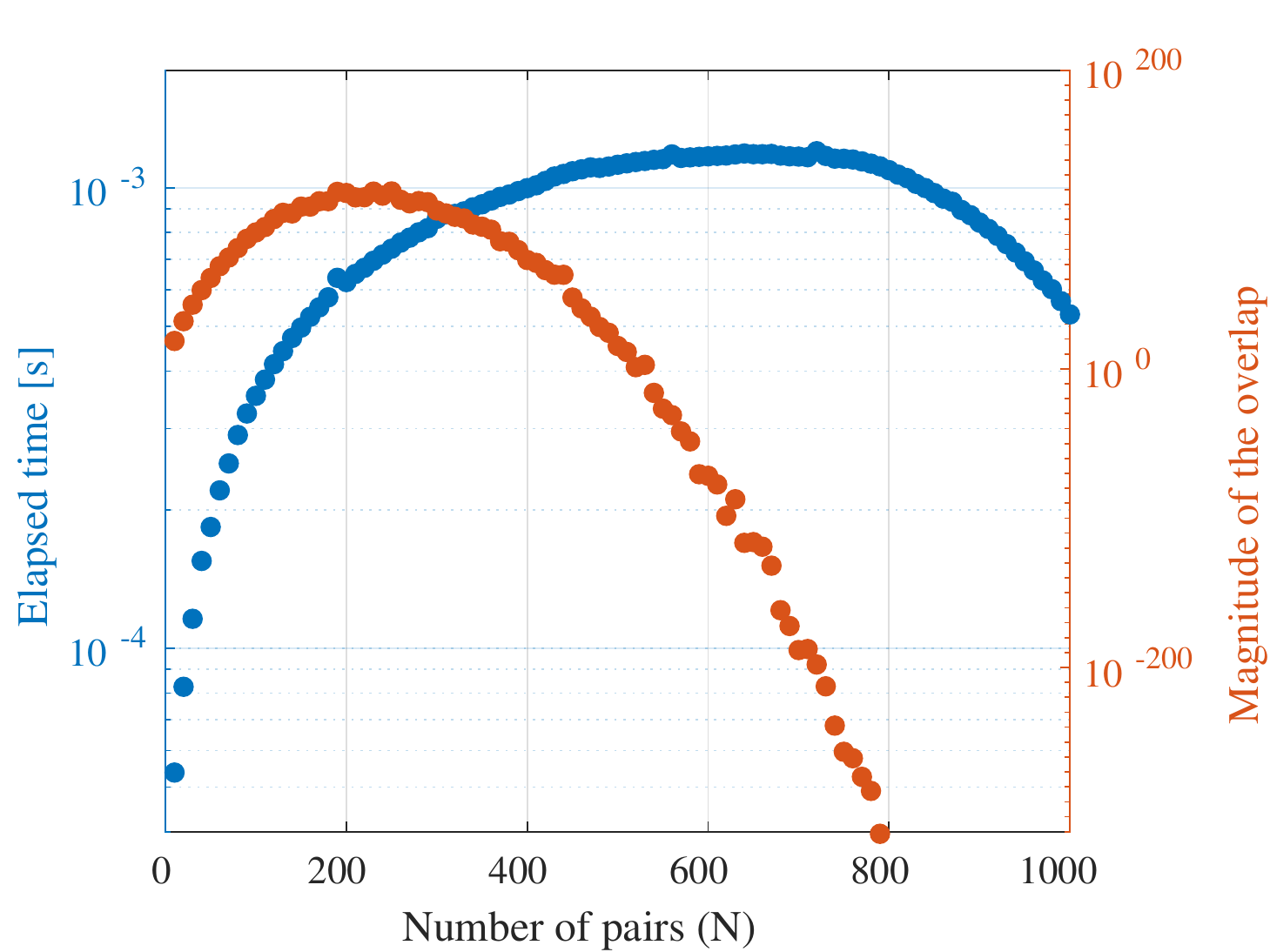}}
        \subfloat[\label{fig:100L_overlap}]
            \centering
            {\includegraphics[width=0.45\textwidth]{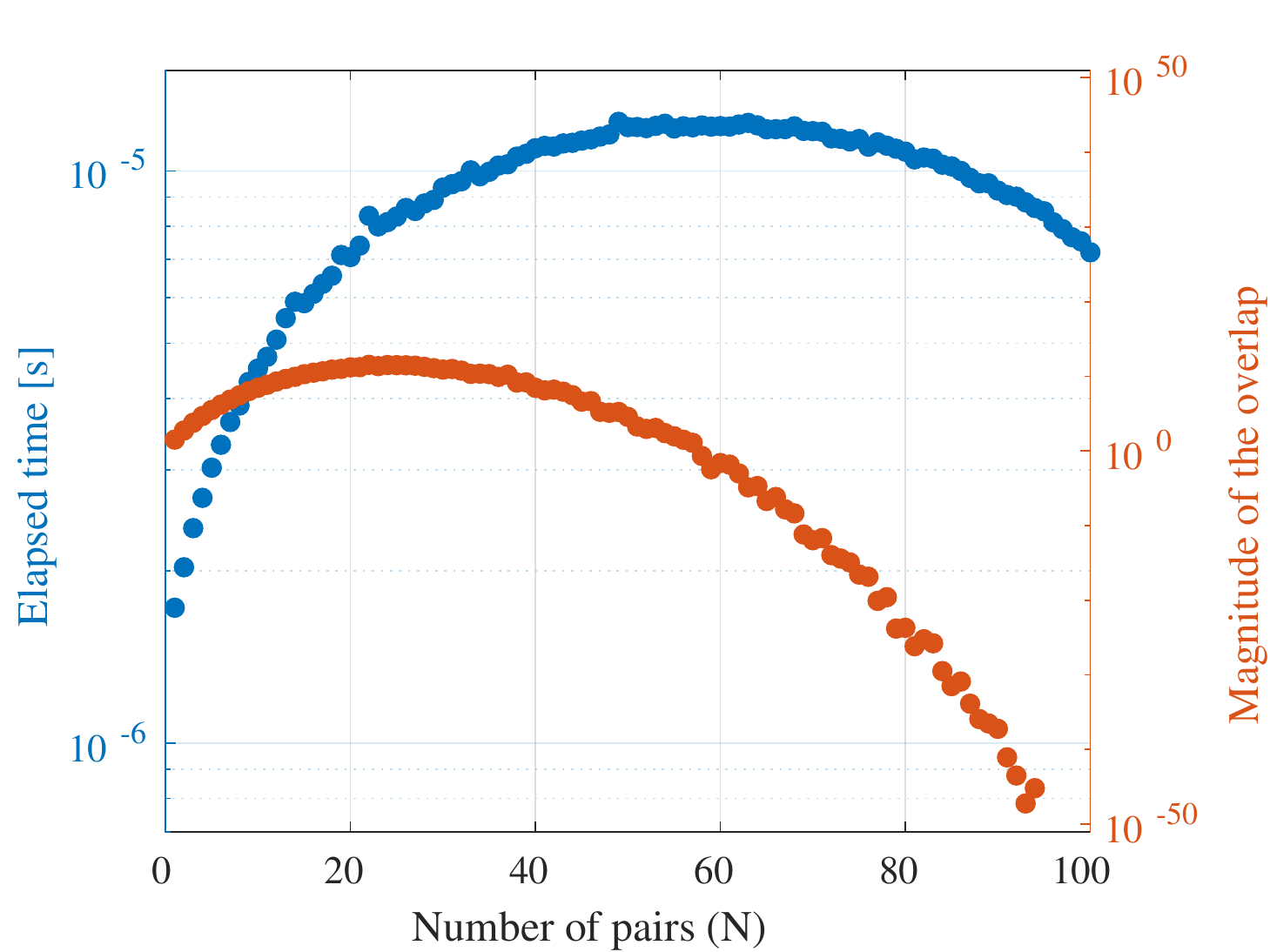}}            
        \caption{Mean of elapsed times (the left axes---blue), and magnitudes of the norm of AGP (the right axes---red) as a function of number of electron-pairs at some fixed number of levels. (a) $M = 1000$ levels, $10 \leq N \leq 1000$ pairs; (b) $M = 100$ levels, $1 \leq N \leq 100$ pairs. Every point is the mean of a random sample with $\eta_p \sim$ Unif(0,1) and $10^3$ observations.} 
        \label{fig:Time&Mag_vs_N}
    \end{figure*}  
    The algorithm we use to calculate the ESP is one initially proposed by Fischer. \cite{gerhard_h._fischer_einfuhrung_1974, baker_computing_1996} This is the same summation algorithm used today in MATLAB's \poly{} function. Following the notation of Ref. \cite{jiang_accurate_2016} we call it \sumESP{}. The \sumESP{} algorithm takes advantage of the following property of ESP
    \begin{equation} \label{eq:ESFprop}
        S_{N}^{M} = S_{N}^{M-1}(\eta_p^2) + \eta_p^2 S_{N-1}^{M-1}(\eta_p^2).
    \end{equation}
    Intuitively, Eq. (\ref{eq:ESFprop}) says that we can split any ESP into two sums such that one contains some arbitrary term $\eta_p^2$ in all of its summands and one that does not. In the context of quantum chemistry, Eq. (\ref{eq:ESFprop}) is mentioned in Ref. \cite{Staroverov_optimization_2002} However, there has been substantial progress in the computational and applied mathematics community to better understand and craft the algorithm. In particular, Ref. \cite{rehman_computing_2011} performed an error analysis and proved the stability of the algorithm.  Ref. \cite{jiang_accurate_2016} made a slight improvement to the roundoff error bound and introduced new ways of calculating \sumESP{} with enhanced precision. From these analyses it follows that \sumESP{} is highly accurate and stable for positive summands \cite{rehman_computing_2011}---as is the case for the overlaps of AGP.
    
    Quantitatively, one can say that, given $\{ \eta_1^2,..., \eta_M^2\}$ as a set of floating-point numbers, the ``worst case" error due to \sumESP{} is bounded above as follows \cite{jiang_accurate_2016}
    \begin{equation}
        \abs{ \frac{S_N^{M} - S_N^{*M}}{S_N^{*M}} } \leq \frac{2(M-1)\epsilon}{1 - 2(M-1)\epsilon},
    \end{equation}
    where $S_N^{*M}$ is the exact value of ESP and $ S_N^{M}$ is computed using \sumESP{} in floating-point arithmetic; $\epsilon$ denotes machine epsilon (unit roundoff). The error bound assumes that there are no numerical overflow and underflow occurring anywhere in the calculation. \cite{rehman_computing_2011} In practice, however, poor choice of normalization of $\eta_p$ could lead to overflow and/or underflow. For example, when $\eta_p^2 > 1$, the sum could overflow even for small values of $M$. Generally, \sumESP{} is better conditioned when $ \eta_p^2 < 1$; this is because the intermediate summations in the algorithm prevents the summands from getting too small. Nevertheless, because the sum is dominated by multiplications in $N > M/2$, one must watch for possible numerical underflow when $M$ is large and $N$ is close to $M$. And when $N < M/2$ the sum is dominated by additions, thus overflow is possible. Fortunately, issues of this kind can be resolved by scaling all geminal coefficients by some constant that prevents overflow or underflow without any hindrance to the method.
    
    To make this analysis more quantitative, for $\eta_p \sim$ Unif(0,1), we have plotted the magnitude of \sumESP{}  (the same as the overlap of AGP) as a function of $M$ in Fig. \ref{fig:overflow_underflow}. In Fig. \ref{fig:Time&Mag_vs_N} (the right axes), we fix $M = 10^2, 10^3$ and vary $N$.
    It follows that in very large systems one must take caution when performing calculations away from $N \ll M$ or half-filling. But in moderate or small systems overflow and underflow should not be a concern. Indeed, the exact regimes in which overflow or underflow is expected highly depends on the distribution of geminal coefficients. Physically, the problem of overflow should be less likely because many of the geminal coefficients approach zero (e.g. near HF limit). And underflow is not of practical concern because in realistic calculations the number of orbitals should be greater than the number of electrons, i.e. $N \ll M$. With our experience using physical geminal coefficients in the pairing Hamiltonian, we have not yet observed overflow or underflow issues.
    
\subsection{Runtime cost of an individual matrix element} \label{sub:IdivnRDM_cost}\
    \begin{figure}[b]
        \centering
        \includegraphics[width=0.42 \textwidth, keepaspectratio]{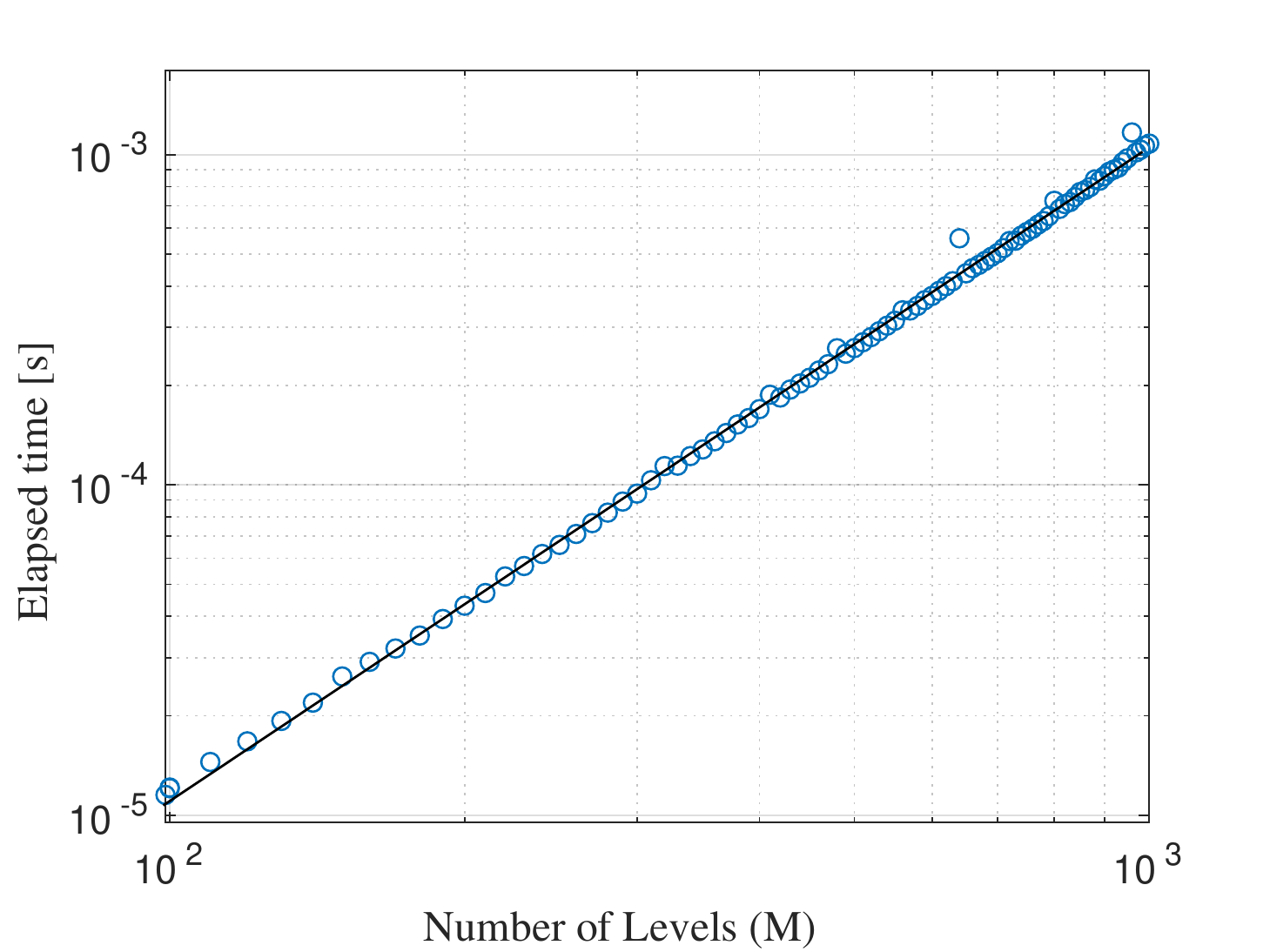}
        \caption{Asymptotic scaling of the norm of AGP at half-filling as a function of system size $M$. A linear fit to the log-log plot shows that the scaling is quadratic.}
        \label{fig:scaling}
    \end{figure}
    The cost of calculating a single matrix element of $n$-pair RDM is bounded above by the cost of the norm of AGP, i.e. $\lAGP AGP \rangle$. This is because all higher order RDMs require evaluation of lower degree ESPs by Eq. (\ref{eq:npRDM}) and the cost of the prefactor is negligible. As such, to get the upper bound of the cost, we only report the elapsed time for evaluating the norm of AGP, which we refer to as the overlap. The theoretical cost of the overlap as a function of $N$ and $M$ grows as $N(1-N + M) - 1$, which is the total number of iterations in the loops of the \sumESP{} algorithm. 
    
    The left axes in Fig. \ref{fig:Time&Mag_vs_N} (the blue dots) illustrate the elapsed time of the overlap as a function of number of pairs, $N$. Every point in the plot is the sample mean of $10^3$ observation points with $\eta_p \sim$ Unif(0,1). By inspection, the most expensive computations occur when $N \approx M/2$ which is expected since ${M \choose N}$ is maximum when $N = M/2$. The shorter elapsed times in $N > M/2$ is due to fewer summations in the algorithm as $N$ approaches $M$. 
    
    To find the asymptotic scaling with system size, $M$, we fix $N = M/2$ and vary $M$ from $10^2$ to $10^3$. This gives the most expensive elapsed time for every value of $M$. The results are shown in Fig \ref{fig:scaling} in which every point is the sample mean of $10^3$ observations with $\eta_p \sim$ Unif(0,1). A linear fit to the log-log plot indicates that the asymptotic time scales quadratically, $T[s] \propto M^{1.97}$, with the system size, in line with the theoretical result.
    
    \subsection{Runtime cost of \textit{n}-pair RDMs} \label{sub:FullnRDM_cost}
    \begin{figure*}[t]
    \centering
    \subfloat[\label{fig:Z02}]
        \centering
        {\includegraphics[width=0.32\textwidth]{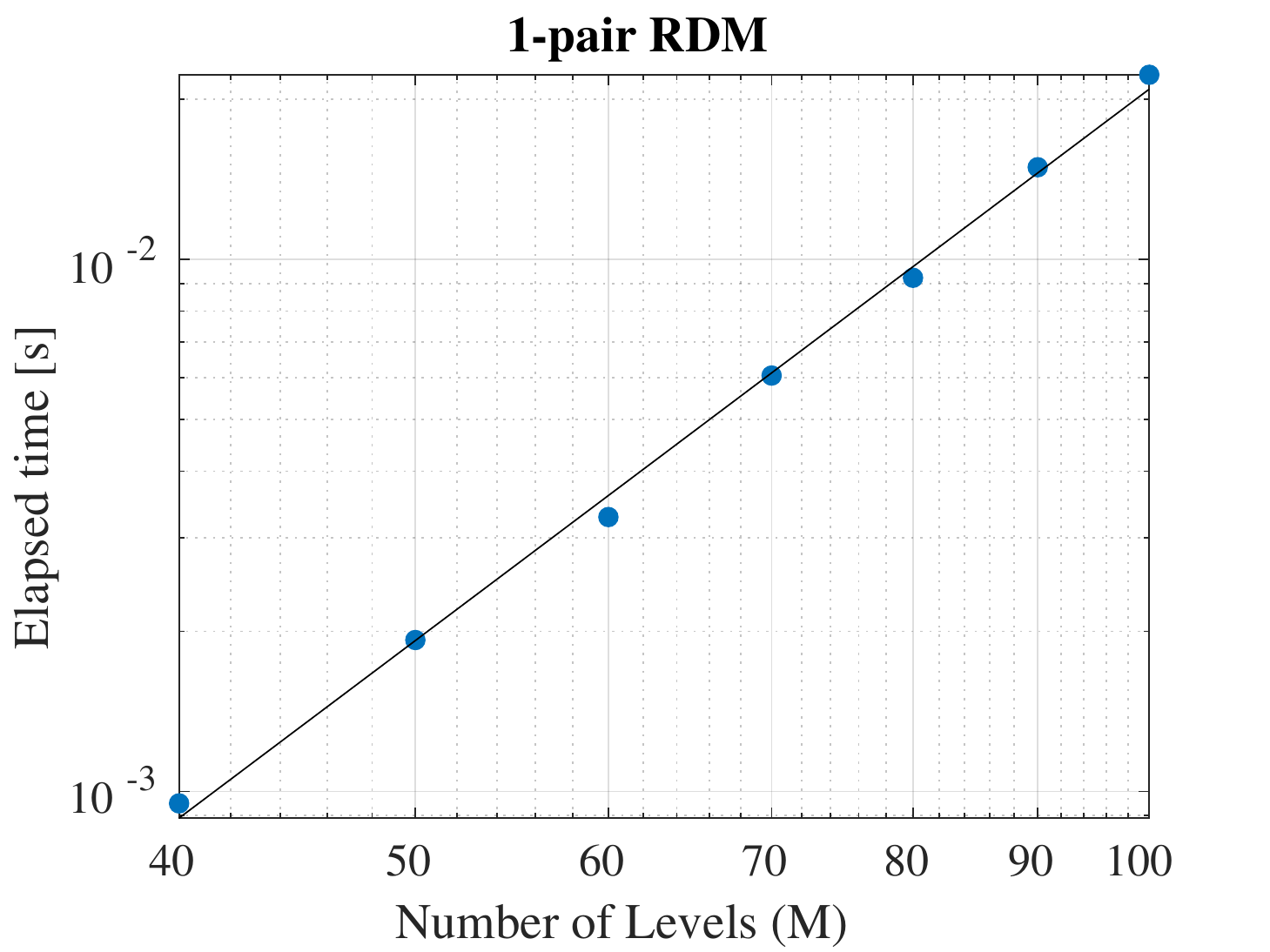}}
    \subfloat[\label{fig:Z04}]
        \centering
        {\includegraphics[width=0.32\textwidth]{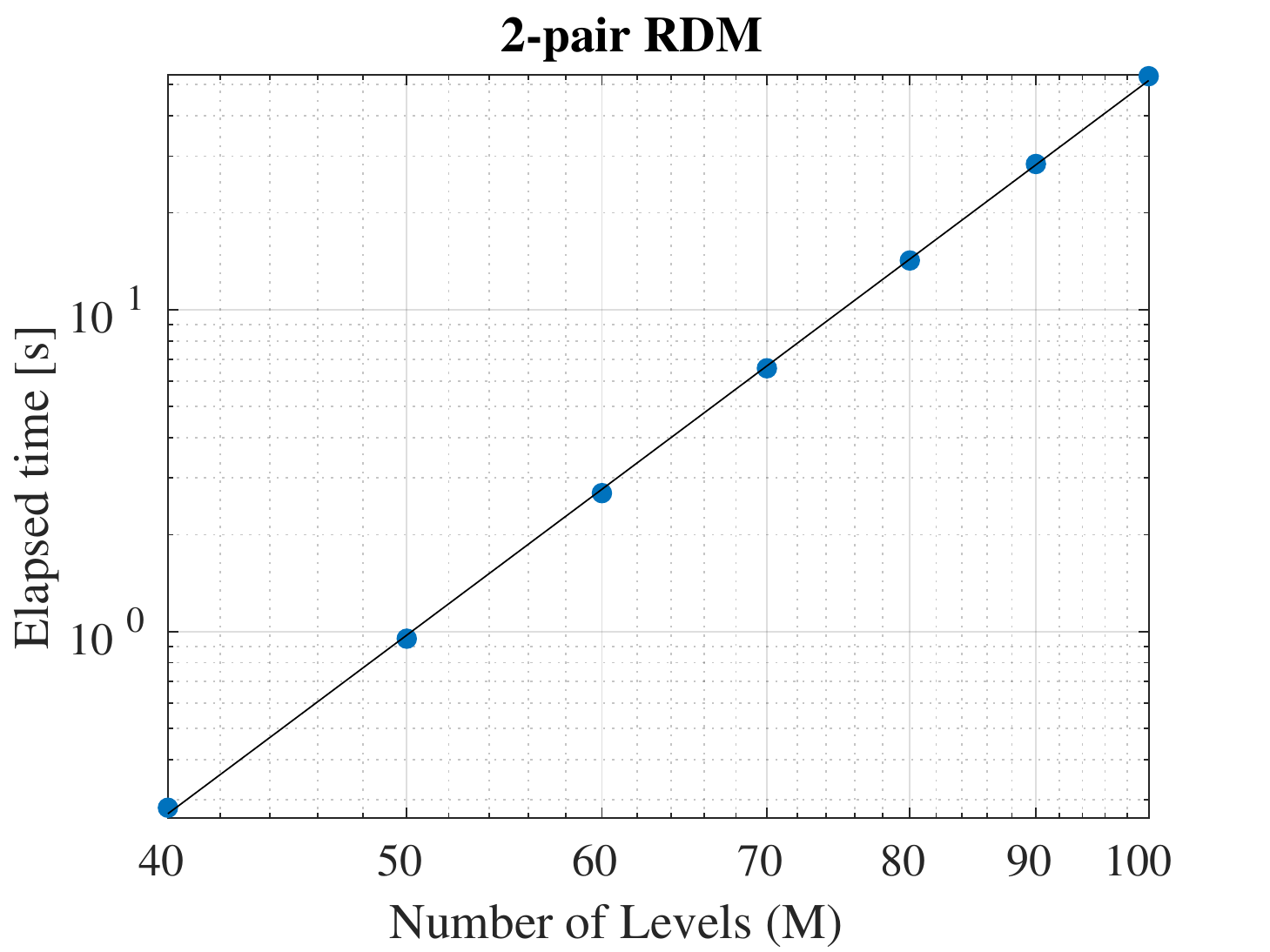}}  
    \subfloat[\label{fig:Z06}]
        \centering
        {\includegraphics[width=0.32\textwidth]{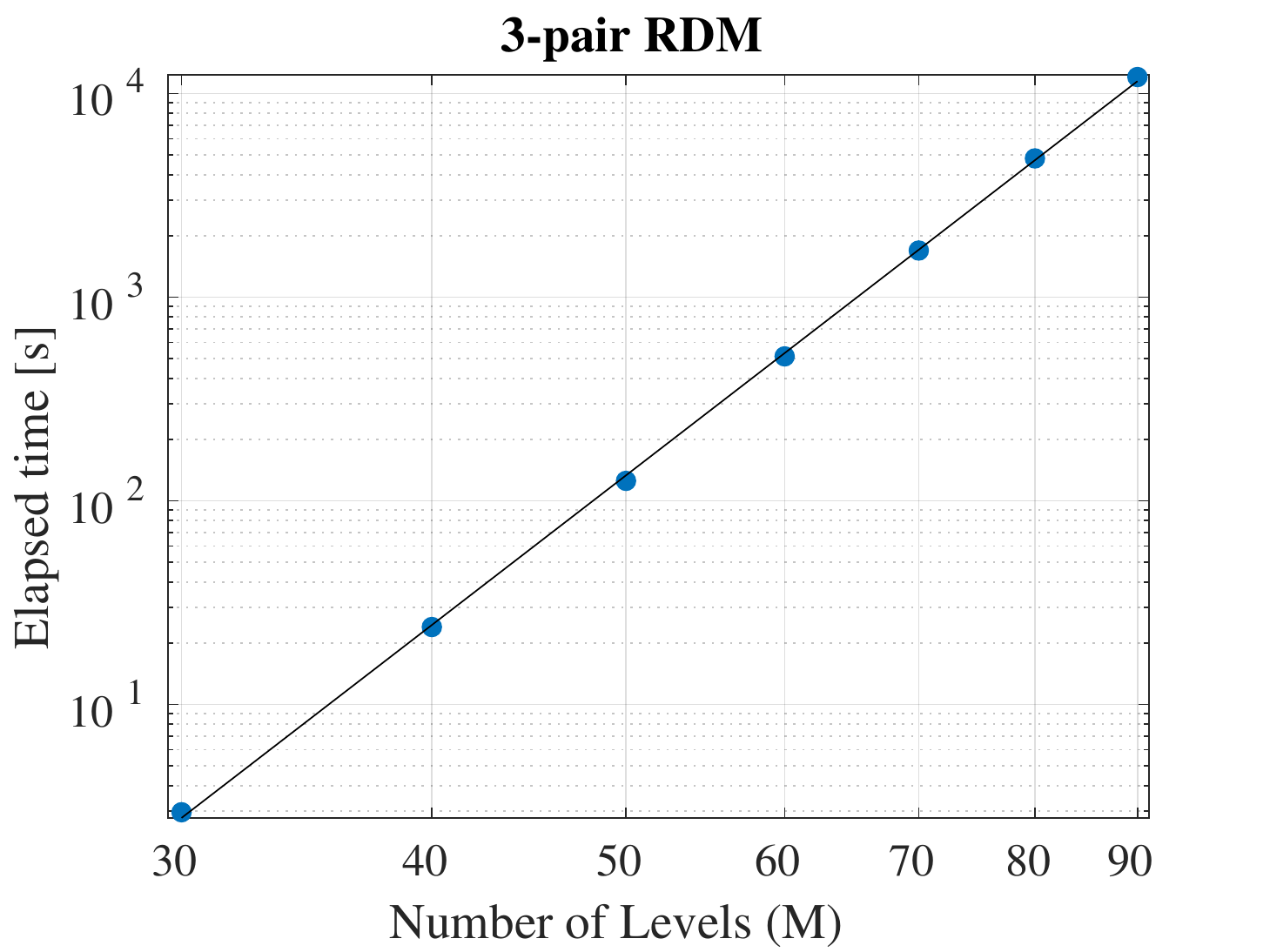}}
    \caption{Time scales of computing $n=1,2, \text{and } 3$ electron-pair RDMs in parallel using Algorithm 1 (see Appendix \ref{sec:appendix1}). The black line is the linear fit to the log-log plot: (a) 1-pair RDM, $T[s] \propto M^{3.4}$; (b) 2-pair RDM, $T[s] \propto M^{5.7}$; (c) 3-pair RDM, $T[s] \propto M^{7.6}$.}
        \label{fig:RDMs}
    \end{figure*}      
    Here, we report the maximum time needed for calculating all matrix elements of $n$-pair RDMs. At this juncture, we remind the reader that it is  sufficient to merely compute and store $\prdm^{p_1<...<p_n}_{q_1<...<q_n}$. Since the calculations of the matrix elements are independent from each other, this is highly parallelizable. (See Appendix \ref{sec:appendix1} for the pseudocode used to calculate the matrix elements.)
    
    The theoretical cost of constructing an $n$-pair RDM on a single core is 
    \begin{equation}
        \texttt{cost} \{\text{ $n$-pair RDM} \} \leq {M \choose n}^2 \texttt{cost} \{ \lAGP AGP \rangle \},
    \end{equation}
    where $ {M \choose n}^2 $ is the number of matrix elements needed to construct an $n$-pair RDM with $M$ levels. Recall that the asymptotic cost of the norm of AGP is $\mathcal{O}(N(M-N)) $. 
    
    In Fig. \ref{fig:RDMs} we report the average elapsed time of $n$-pair RDM for $n = 1,2, \text{and } 3$, with $M \leq 100$. The plots show the sample average of 100 observations for $n = 1,2$ and $10$ observations for $n = 3$ in which $\eta_p \sim$ Unif(0,1). 

\section{Reconstruction formulae} \label{sec:reconstruction}

In Sec. \ref{sub:FullnRDM_cost} we argued that the asymptotic cost of constructing an $n$-pair RDM is $\mathcal{O}(NM^{2n+1})$. Here, we show a way of cutting down the cost by expressing higher order RDMs as a linear combination of lower order ones and geminal coefficients. This is an important ingredient for developing efficient correlated theories based on AGP, as most techniques require evaluation of high order RDMs. For example, in AGP based configuration interaction (AGP-CI) calculations done in our group \cite{henderson_geminal-based_2019} up to to 5-body density matrices were needed (the Hamiltonian and the correlators contain 2-body operators each, and using the killers one can introduce a commutator to reduce the rank by 1). Therefore, a systematic way of reducing the scaling and the cost of calculating many-body RDMs is of great interest. Rosina mathematically anticipated \cite{cioslowski_many-electron_2000} that the 2-RDMs of AGP determine all of its higher order RDMs. As it turns out, we here show that 1-RDM occupation number RDMs ($\nu^{(1)}$) and $\eta$'s are sufficient to determine all RDMs of AGP. 

Note that the decomposition method presented do not reflect vanishing cumulant decomposition of AGP density matrices. Also note that, the method differs from the decomposition of PBCS in which the density matrix is a weighted integral over the projection grid of a factorized transition density matrix. \cite{scuseria_projected_2011}

\subsection{Direct decomposition}
Our goal, here, is to express the nonzero elements of any $n$-pair RDM, $\prdm^{(n)}$, as a linear combination of ${\nrdm}^{(n')}$ where $ n \leq n' \leq 2n$; then we show that we can further decompose each $\nrdm^{(n')}$ into a sum of $\nrdm^{(1)}$, hence  $\prdm^{(n)} \xrightarrow{} \sum \nrdm^{(1)}$. 

Consider a nonzero element of an $n$-pair RDM, $\prdm^{p_1...p_n}_{q_1...q_n}$. In general, it can be that $p_i = q_j$ for some $i,j$. Since $\N_p = 2\Pdag_p \P_p$, we can write
\begin{align} \label{eq:Gamm_diff_indices}
    \begin{split}
        \prdm^{p_1...p_n}_{q_1...q_n} = {2^{-k}} \nonumber 
    \end{split} \\
    \begin{split}
        \lAGP \Pdag_{p_1}...\Pdag_{p_{n-k}} \N_{r_1}...\N_{r_k} \P_{q_1}...\P_{q_{n-k}} \rAGP,
    \end{split}    
\end{align}
where $k$ is the number of common indices among $\{p_i\}$ and $\{q_i\}$. Written in this manner, we can assume all the remaining indices are different; otherwise the element is zero by construction. By this and hermiticity of $\prdm^{(n)}$, and the fact that the top and lower indices are permutable, we can further express 
\begin{align}\label{eq:Gamm_preKiller}
    \begin{split}
        \prdm^{p_1...p_n}_{q_1...q_n} = \frac{1}{2^{n}} \lAGP \N_{r_1}...\N_{r_k} (\Pdag_{p_1} \P_{q_1} + \Pdag_{q_1} \P_{p_1}) \nonumber
    \end{split} \\
    \begin{split}
        ...(\Pdag_{p_{n-k}} \P_{q_{n-k}} + \Pdag_{q_{n-k}} \P_{p_{n-k}}) \rAGP.
    \end{split}      
\end{align}
Now, by manipulating the killer of AGP, i.e. $\K_{pq} \rAGP = 0$ reported in Ref. \cite{henderson_geminal-based_2019} we can write
\begin{align}\label{eq:Killer}
    \begin{split}
        \Pdag_p \P_q + \Pdag_q \P_p = \nonumber 
    \end{split} \\
    \begin{split}
        \frac{1}{\eta_p^2 + \eta_q^2} \left( \K_{pq}^{\dagger} + \K_{pq} + \eta_p\eta_q (\N_p + \N_q - \N_p\N_q ) \right).
    \end{split}    
\end{align}
By plugging Eq. (\ref{eq:Killer}) into Eq. (\ref{eq:Gamm_preKiller}) we arrive at an expression for $\prdm^{(n)}$ that is written purely as a linear combination of $\nrdm^{(n')}$ and geminal coefficients as desired. Now, we decompose $\nrdm^{(n')}$ into a sum of $\nrdm^{(1)}$ and factors of $\eta_p$. For this, we introduce a closed form expression which we prove in Appendix \ref{sec:proof}
\begin{equation}\label{eq:Direct_breaksown}
    \nrdm_{p_1 ... p_n} = \sum_{i=1}^{n} \left( \displaystyle\prod_{\substack{j \geq 1 \\ j \neq i}}^{n} \frac{2\eta_{p_j}^2}{ \eta_{p_j}^2 - \eta_{p_i}^2} \right) \nrdm_{p_i} .
\end{equation}

To make this discussion concrete, consider $\prdm^{(2)}$ as an example, whose non-zero elements can be written as 
\begin{equation}
    \prdm^{pq}_{rs} =
    \begin{cases*}
      \langle \Pdag_{p}\Pdag_{q} \P_{r} \P_{s} \rangle & if $p,q \neq r,s$ \\
      \frac{1}{2} \langle \Pdag_{p} \N_{q} \P_{r} \rangle & if $p \neq r$ and $q = s$\\
      \frac{1}{4} \langle \N_{p} \N_{q} \rangle & if $p = r$ and $q = s$ \\
    \end{cases*}
\end{equation}
For each of the three cases, we obtain an expression entirely in terms of the number RDMs and $\eta$'s using Eq. (\ref{eq:Killer}):
\begin{equation}
    \prdm^{pq}_{rs} =
    \begin{cases*}
      \frac{\eta_p \eta_q \eta_r \eta_s}{4 (\eta_p^2 + \eta_r^2)(\eta_q^2 + \eta_s^2)} \big\langle (\N_p + \N_r - \N_p \N_r) \\
      (\N_q + \N_s - \N_q \N_s) \big \rangle\\ 
      \\
      \frac{\eta_p\eta_r}{4 (\eta_p^2 + \eta_r^2)} \big( \langle \N_{p} \N_{q} \rangle + \langle \N_{q} \N_{r}\rangle - \langle \N_{p} \N_{q} \N_{r}\rangle \big) \\
      \\
      \frac{1}{4} \langle \N_{p} \N_{q} \rangle
    \end{cases*}
\end{equation}
Then each of the $\langle \N_p \N_q \rangle$, $\langle \N_p \N_q \N_r \rangle$, etc., can be inserted in Eq. (\ref{eq:Direct_breaksown}) to produce an expression in terms of a linear combination of $\langle {\N_p} \rangle$ and $\eta$'s only, as desired.

This result is profound as it implies that we can obtain all higher rank pair RDMs by merely computing $\nrdm^{(1)}$ whose cost grows asymptotically as $\mathcal{O}(M^{2}N)$ and may be computed only once for the rest of the calculations. The cost of prefactors is $n(n-1)$ which is negligible  since for all practical purposes $n \ll N, M$. However, we pay the price of introducing  $1/(\eta_p^2 - \eta_q^2)$ factors that can be numerically ill-posed when $\eta_p^2 \approx \eta_q^2$ or when $\eta_p,\eta_q \xrightarrow{} 0$. In the regime that these factors are not problematic, the decomposition $\prdm^{(n)} \xrightarrow{} \sum \nrdm^{(1)}$ is a major improvement over computing all the matrix elements of an $n$-pair RDM. We must note that in our own implementation of these equations for practical problems (e.g. the attractive pairing Hamiltonian in Ref.\cite{henderson_geminal-based_2019}) we have not observed these potential numerical issues. 

\subsection{Stepwise decomposition}
In practice, there are situations in which it is more advantageous to break down high rank RDMs in terms of ``slightly" lower rank ones. This also makes the issue of having too many $1/(\eta_p^2 - \eta_q^2)$ factors less severe. To this end, we need a new notation for $\langle \Pdag...\N...\P...\rangle$. Define
\begin{align}
    \begin{split}
        \mathlarger{Z}^{(k,m)}_{p_1...p_t, \ r_1...r_k, \ q_1...p_t} = \nonumber
    \end{split} \\
    \begin{split}
         \lAGP \Pdag_{p_1}...\Pdag_{p_t} \N_{r_1}...\N_{r_k} \P_{q_1}...\P_{q_t}\rAGP,
    \end{split}     
\end{align}
where $k$ is the number of $\N_p$ operators in the middle, and $m$ is the total number of indices such that $k \leq m$. And define $t = {(m-k)}/{2}$. For example, 
\begin{eqnarray}
    &{}&\z^{(0,6)}_{pqr,stu} = \lAGP \Pdag_p \Pdag_q \Pdag_r \P_s \P_t \P_u\rAGP \nonumber\\
    &{}&\z^{(1,5)}_{pq,r,st} = \lAGP \Pdag_p \Pdag_q \N_r \P_s \P_t \rAGP \nonumber\\
    &{}&\z^{(2,4)}_{p,qr,s} = \lAGP \Pdag_p \N_q \N_r \P_s \rAGP \nonumber\\
    &{}&\z^{(3,3)}_{pqr} = \lAGP \N_p \N_q \N_r \rAGP.  \nonumber 
\end{eqnarray}
Notice that the subscript indices on each $Z^{(k,m)}$ are all different; otherwise they are either zero or reducible to some other $Z^{(k',m')}$ by $\N_p = 2\Pdag_p \P_p$ and $\N_p^2 = 2\N_p$. Obviously all $\prdm^{(n)}$ can be mapped to $Z^{(k,2n-k)}$ for some $k$ and $n$, and vice versa. 

Our goal, here, is to show that 
\begin{eqnarray}
    Z^{(0,2n)} &\xrightarrow{}& \sum Z^{(1,2n-1)} \xrightarrow{}...\xrightarrow{} \; \sum Z^{(n,n)} = \sum \nrdm^{(n)} \nonumber\\
    &\xrightarrow{}& \sum \nrdm^{(n-1)}  \xrightarrow{} .... \xrightarrow{} \sum \nrdm^{(1)}.
\end{eqnarray}
As we prove in Appendix \ref{sec:proof}, this can be accomplished by using the following formula at every step
\begin{eqnarray}\label{eq:stepWise_breaksown}
    \z^{(k,m)}_{p_1...p_{t}, \ r_1...r_k, \ q_1...q_{t}} = \frac{\eta_{p_i}\eta_{q_j}}{ 2(\eta_{q_j}^2 - \eta_{p_i}^2) } \big( \nonumber \\
     \z^{(k+1,m-1)}_{...p_{i-1}p_{i+1}..., \ r_1...r_k q_{j}, \ ...q_{j-1}q_{j+1}...} - \nonumber \\
     \z^{(k+1,m-1)}_{...p_{i-1}p_{i+1}..., \ p_i r_1...r_k, \ ...q_{j-1}q_{j+1}...} \big),
\end{eqnarray}
where $i,j \in \{1,2,...t \}$. Similarly, it is easy to show that \begin{eqnarray}\label{eq:stepWise_breaksown_nu}
    \nrdm_{p_1 ... p_{n}} &=& \frac{ 2 \eta_{p_{j}}^2 }{\eta_{p_{j}}^2 - \eta_{p_i}^2} \nrdm_{p_1 ...p_{j-1}p_{j+1}...p_{n}} \nonumber \\
    &+& \frac{ 2 \eta_{p_i}^2}{ \eta_{p_i}^2 - \eta_{p_{j}}^2 } \nrdm_{p_1 ...p_{i-1}p_{i+1}...p_{n}} ,
\end{eqnarray}
here, $i,j \in \{1,2,...n\}$. 

The advantage of breaking down the density matrices like this is that, at every step, we reduce the dimension by one, thereby reducing the asymptotic scaling of computing it by a factor of $M^2$ and a negligible prefactor. Moreover, we can stop at any step of our choice based on the cost that we are willing to tolerate. 

\section{Runtime of Energy and AGP-CI} \label{sec:Energy}
    \begin{figure}[t]
        \centering
        \includegraphics[width=0.45 \textwidth, keepaspectratio]{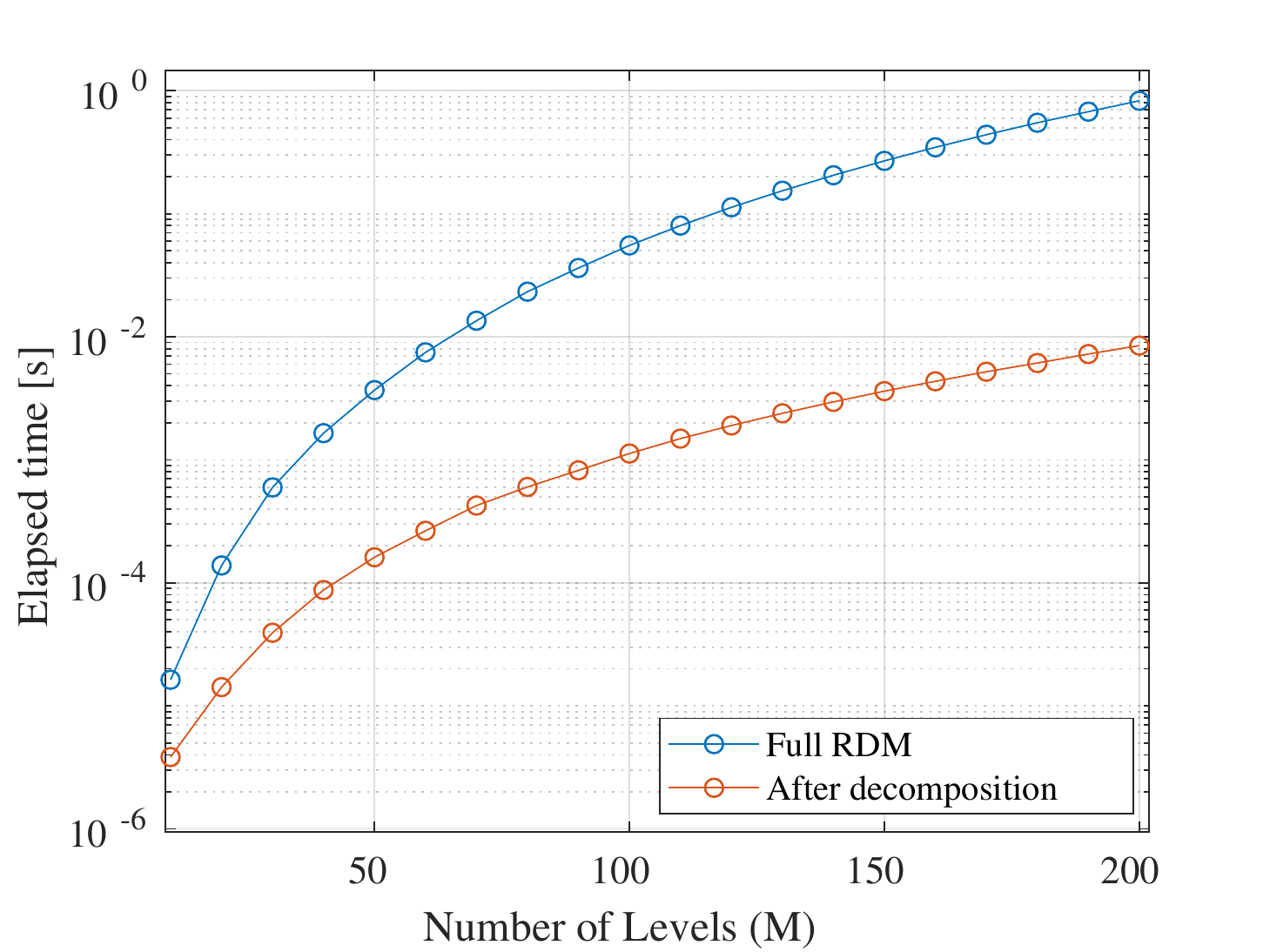}
        \caption{Runtime measurement of the pairing Hamiltonian energy with and without the reconstruction formula. Each point is the sample mean of 100 observations with $\eta_p \sim \text{Unif}(0,1)$.}
        \label{fig:energy}
    \end{figure}
    
In this section, we benchmark the speed gain when using the reconstruction formulae. Here, we report the runtime measurements of two calculations: (1) the energy of the pairing Hamiltonian (reduced BCS); (2) AGP-CI calculations as reported in Ref. \cite{henderson_geminal-based_2019}. See Appendix \ref{sec:Computer_env} for the computer environments used in these benchmarks.

We start with the pairing Hamiltonian. Recall that the attractive pairing Hamiltonian can be written as follows: \cite{degroote_polynomial_2016}
\begin{equation}
    \mathlarger{H} = \sum_p \epsilon_p \N_p - G \sum_{pq} \Pdag_p \P_q. 
\end{equation}
The expected value of the energy over AGP in terms of the pair RDMs is
\begin{equation}  \label{eq:Energy}
    E = \sum_p  \prdm^{p}_{p} (2\epsilon_p - G)  - 2G \sum_{p<q} \prdm^{p}_{q}.
\end{equation}
And using the reconstruction formulae, we can get
\begin{equation} \label{eq:Energy_decomposed}
    E = \sum_p  \nrdm_{p} (\epsilon_p - \frac{G}{2})  - G \sum_{p<q} \frac{\eta_p\eta_q}{\eta_p^2 - \eta_q^2} (\nrdm_{p} - \nrdm_{q}).
\end{equation}

Fig. \ref{fig:energy} shows the elapsed time differences between implementing Eq. (\ref{eq:Energy}) and Eq. (\ref{eq:Energy_decomposed}) as a function of number of levels at half-filling. For each case, the corresponding RDM is stored and then called in the calculation. By inspection, it is easy to see that the reconstruction formulae speed up the calculations often by an order of magnitude. Obviously, the improvement becomes more noticeable as the number of levels increases. 

\begin{figure}[t]
    \centering
    \includegraphics[width=0.45 \textwidth, keepaspectratio]{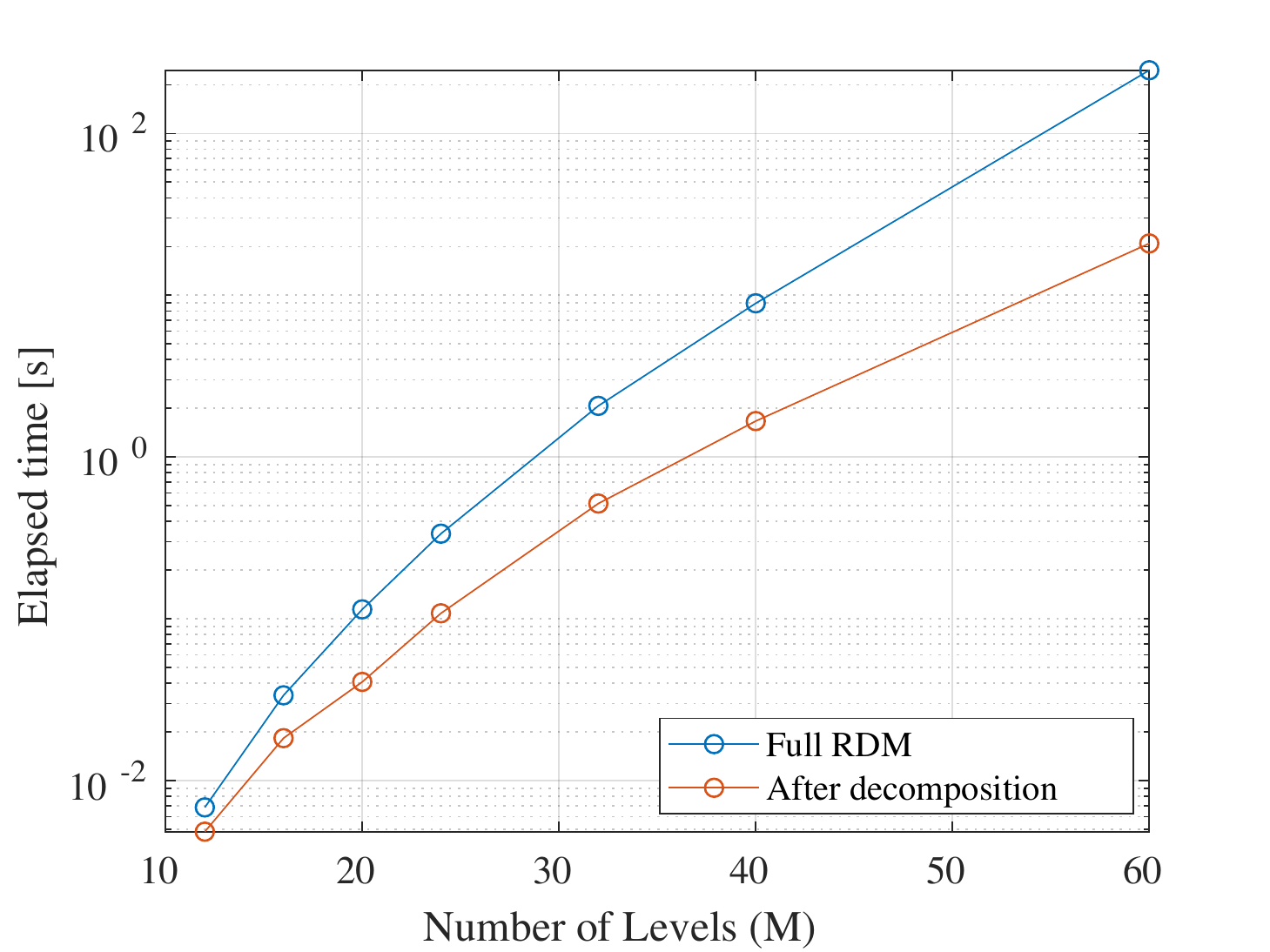}
    \caption{Runtime cost of AGP-CI calculation with and without the reconstruction formula. Every point is the mean of the same calculation repeated $200$ times.}
    \label{fig:AGPCI}
\end{figure}

At last, we report the elapsed time of performing AGP-CI calculations with and without the reconstruction formulae in Fig. \ref{fig:AGPCI}. For these calculations, we stored up to $4$-indexed RDMs in memory and calculated all the rank 5 RDMs using the direct decomposition formula Eq. (\ref{eq:Direct_breaksown}). Here, $\eta$'s are optimized beforehand for the pairing Hamiltonian ($G=2$) for various values of $M$. Similar to the energy calculations, we observe that the reconstruction formulae lead to a substantial improvement. 

\section{Conclusions} \label{sec:conclusions}

Analytic expressions of the norm and reduced density matrices of AGP wavefunction are proportional to ESP. We used the \sumESP{} algorithm \cite{rehman_computing_2011, jiang_accurate_2016} to efficiently calculate ESPs and argued that, with appropriate normalization, it is well suited for physical problems wherein $N \ll M$. We have shown that our method can reliably calculate the norm and elements of RDMs to all ranks for systems as large as a few thousand orbitals and hundreds of electrons. Our runtime measurements indicate that the asymptotic cost per element of an $n$-pair RDMs is at most quadratic, and the cost of building an $n$-pair RDM grows asymptotically as $\mathcal{O}(NM^{2n+1})$. 
    
However, to reduce the cost of computing the RDMs even further, we have derived reconstruction formulae that allow decomposition of any RDM into a linear combinations of lower rank ones and geminal coefficients. We introduced two methods: (1) Direct decomposition---breaks down a high rank pair RDM in terms of linear combination of rank-1 occupation RDMs; (2) Stepwise decomposition---reduces the dimension of a RDM by one at every step thereby reducing the cost of computing it by a factor of $M^2$.
    
We demonstrate the advantage of using our reconstruction formulae by benchmarking it against the energy of the pairing Hamiltonian and AGP-CI calculation without the reconstruction formulae. The numerical results indicate that, indeed, the reconstruction formulae lead to a substantial speed-up, especially in systems with large number of orbitals. 

\section*{Acknowledgements} \label{sec:acknowledgements}
This work was supported by the U.S. National Science Foundation under Grant No. CHE-1762320; and by the Big-Data Private-Cloud Research Cyberinfrastructure MRI-award funded by NSF under grant CNS-1338099 and by Rice University. G.E.S. is a Welch Foundation Chair (Grant No. C-0036). 

\appendix

\section{Numerical Implementation} \label{sec:appendix1}

Below is a pseudocode for calculating a single matrix element of an $n$-pair RDM. Implementation of \sumESP{} is taken from Ref. \cite{jiang_accurate_2016}
    
    \begin{algorithm}[H]
    \SetAlgoLined
    \SetKwData{Left}{left}\SetKwData{This}{this}\SetKwData{Up}{up}
    \SetKwFunction{Union}{Union}\SetKwFunction{FindCompress}{FindCompress}
    \SetKwInOut{Input}{input}\SetKwInOut{Output}{output}
    \Input {$ \mathbf{x} = \{ \eta_1,..,\eta_M \} $; \\ 
            $N$: number of pairs; \\
            $L = \{ p_1,...p_n,q_1,...,q_n \}$ }
    \Output{ $\prdm^{p_1...p_n}_{q_1,...q_n} $ } \
    
    \textbf{integer} $i,j,k, n, N, M$; \quad \textbf{real} $p$  
    
    \textbf{integer array} $(L_i)_{1:2n}$ 
    
    \textbf{real array} $(L_i)_{1:2n}$, $(x_i)_{1:M}$, $(S_j^{i})_{( i\in 1:M)( j\in 1:N+1)}$ 

    \BlankLine
     \If { $(p_i = p_j$ or $q_i = q_j)$ or $(n > N)$ } {
        Output = 0; 
        
        End the program;
     }
     
     $k \xleftarrow{} N - size(L)/2$;   // $size(x)$ gives the size of $x$
    
     $p \xleftarrow{} (\prod_{i \in L} x_i $); 
     
     $x \xleftarrow{} eliminate(x_{i \in L})$ // removes $x_{i \in L}$ from $x$
     
      $x \xleftarrow{} x^2$ 
      
      \BlankLine
     // Below  is the \sumESP{} algorithm
     
     $M \xleftarrow{} size(x)$;
     
     $ S_{j}^i \xleftarrow{}$ zeros for $j>i$;
     $S_0 ^{i} \xleftarrow{} 1$ for $ 1 \leq i \leq M-1$;  
     
     $S_1 ^{1} \xleftarrow{} x_1;$
     
     \For{i = 2 to M}{
      \For{ $j = max \{ 1, i+k-n \}$ to $min \{ i, k \}$ } {
        $S_j^i = S_j^{i-1} + x_i S_{j-1}^{i-1}$
      }
     }
     // End of \sumESP{} \\
     \BlankLine
     
     Output $\xleftarrow{} p \times S_{k+1}^M$
     \caption{Matrix elements of $n$-pair RDM }
     \end{algorithm}
    
One can store any RDM as a 2 dimensional array by linear indexing $p_i$'s and $q_i$'s such that $\prdm^{p_1<...<p_n}_{q_1<...<q_n} \rightarrow \prdm^{\mu}_{\nu}$. The reader may find the following relation handy in the parallel implementation 
\begin{eqnarray}
    \mu &=&  p_1 + \sum_{i=2}^{n} {p_i - 1 \choose i} \nonumber; \quad 1 \leq p_1<...<p_n \leq M.
\end{eqnarray}

\section{Proofs of the reconstruction formulae} \label{sec:proof}

We set out by first deriving Eq. (\ref{eq:stepWise_breaksown}). Notice, by Eq. (\ref{eq:npRDM}) and Eq. (\ref{eq:ESFprop}), we can write
\begin{align} \label{Zk+1_q}
    \begin{split}
        \z^{(k+1,m-1)}_{p_1...p_{i-1}p_{i+1}...p_{t}, \ r_1...r_k q_{j}, \ q_1...q_{j-1}q_{j+1}...q_{t}} =  2^{k+1}\nonumber    
    \end{split} \\
    \begin{split}
         (\eta_{q_j}^2 \displaystyle\prod_{s=1}^{k} \eta_{r_s}^2) (\displaystyle\prod_{ s \neq i,j }^{t}\eta_{p_s} \eta_{q_s}) S_{N-t-k}^{M - m + 1} (\eta_{p_1}^2...\eta_{p_{i-1}}^2\eta_{p_{i+1}}^2...\eta_{q_{t}}^2) 
        \nonumber    
    \end{split} \\        
    \begin{split}
         =  2^{k+1} (\eta_{q_j}^2 \displaystyle\prod_{s=1}^{k} \eta_{r_s}^2) (\displaystyle\prod_{ s \neq i,j }^{t}\eta_{p_s} \eta_{q_s}) \Big( S_{N-t-k}^{M-m} (\eta_{p_1}^2...\eta_{q_{t}}^2) \ + 
        \nonumber    
    \end{split} \\    
    \begin{split}
        \qquad \eta_{p_i}^2 S_{N-t-k-1}^{M-m} (\eta_{p_1}^2...\eta_{q_{t}}^2) \Big) \nonumber        
    \end{split} \\    
    \begin{split}
        =         2 \frac{\eta_{q_j} }{ \eta_{p_i} } \z^{(k,m)}_{p_1...p_{t}, \ r_1...r_k,\ q_1,...q_{t}} + 
    \end{split} \\
    \begin{split}
        \qquad
        \frac{1}{2} \z^{(k+2,m)}_{...p_{i-1}p_{i+1}..., \ p_i r_1...r_k q_j, \ ...q_{j-1}q_{j+1}...} \nonumber
    \end{split}    
\end{align}
Similarly, we can state that
\begin{eqnarray} \label{Zk+1_p}
    &{}&\z^{(k+1,m-1)}_{p_1...p_{i-1}p_{i+1}...p_t, \ p_{i}r_1...r_k ,q_1 \ ...q_{i-1}q_{i+1}...q_t} = \nonumber \\  
    &{}& \qquad 2 \frac{\eta_{p_i} }{ \eta_{q_j} } \z^{(k,m)}_{p_1...p_{t}, \ r_1...r_k,\ q_1,...q_{t}} + \nonumber \\
    &{}& \qquad \frac{1}{2} \z^{(k+2,m)}_{...p_{i-1}p_{i+1}..., \ p_i r_1...r_k q_j, \ ...q_{j-1}q_{j+1}...}.     
\end{eqnarray}
By subtracting Eq. (\ref{Zk+1_p}) from Eq. (\ref{Zk+1_q}) and rearranging the terms we get Eq. (\ref{eq:stepWise_breaksown}) as desired. \qed

Now we prove Eq. (\ref{eq:Direct_breaksown}) by induction on $n$. For $n=1$, the equality is trivially true. Now, given the induction hypothesis for $n$, we want to show the $n+1$ case. Similar to the derivation above, by using Eq. (\ref{eq:npRDM}) and Eq. (\ref{eq:ESFprop}), we can write
\begin{eqnarray*}
    \nrdm_{p_1...p_n} &=& \frac{2 \eta_{p_1}}{\eta_{p_{n+1}}} \z^{(n-1,n+1)}_{p_1, \ p_2...p_n, \ p_{n+1}} + \frac{1}{2} \nrdm_{p_1...p_{n+1}}. \\
\end{eqnarray*}
Using Eq. (\ref{eq:stepWise_breaksown}) and rearranging the terms we get
\begin{eqnarray*}
    \nrdm_{p_1 ... p_{n+1}} &=& \frac{ 2 \eta_{p_{n+1}}^2 }{\eta_{p_{n+1}}^2 - \eta_{p_1}^2} \nrdm_{p_1 ... p_n} \\
    &+& \frac{ 2 \eta_{p_1}^2}{ \eta_{p_1}^2 - \eta_{p_{n+1}}^2 } \nrdm_{p_2 ... p_{n+1}}.
\end{eqnarray*}

Now, we apply the induction hypothesis to $\nrdm_{p_1...p_n}$ and $\nrdm_{p_2...p_{n+1}}$. Brute force algebra shows that
\begin{eqnarray*}
    \nrdm_{p_1 ... p_{n+1}} = (\prod_{j=2}^{n+1} \frac{2\eta_{p_j}^2}{\eta_{p_j}^2 - \eta_{p_1}^2}) \nrdm_{p_{1}} + 
    (\prod_{j=1}^{n}\frac{2\eta_{p_j}^2}{\eta_{p_j}^2 - \eta_{p_{n+1}}^2}) \nrdm_{p_{n+1}} \\
    + \sum_{i=2}^{n} (\prod_{ \substack{j \geq 2 \\j \neq i} }^{n} \frac{ 2\eta_{p_j}^2 }{\eta_{p_j}^2 - \eta_{p_i}^2} )(
    \frac{2 \eta_{p_{n+1}}^2}{\eta_{p_{n+1}}^2 - \eta_{p_i}^2}  \frac{ 2 \eta_{p_{1}}^2}{\eta_{p_{1}}^2 - \eta_{p_i}^2} ) \nrdm_{p_i} \\
    = \sum_{i=1}^{n+1} \left( \displaystyle\prod_{\substack{j \geq 1 \\j \neq i} }^{n+1} \frac{2\eta_{p_j}^2}{ \eta_{p_j}^2 - \eta_{p_i}^2} \right) \nrdm_{p_i}.
    \qed
\end{eqnarray*}

\section{Computer environments} \label{sec:Computer_env}

Here we detail the computer environments used for each runtime test.

Sec. \ref{sub:IdivnRDM_cost} and the energy calculations in Sec. \ref{sec:Energy}: single core of a workstation with Intel Xeon(R) CPU E3-1270 v6, with 8 cores, each at 3.80GHz on a x86\_64 hardware architecture with GNU/Linux operating system. The programs were compiled using GNU Fortran (GCC) 4.8.5 (Red Hat 4.8.5-36) with the default compiler optimization options. 

Sec. \ref{sub:FullnRDM_cost}: the computations were carried out in parallel using 16 cores on a single node of a cluster running on the Rice Big Research Data (BiRD) cloud infrastructure. The environment is as follows: Intel(R) Xeon(R) CPU E5-2650 v2 @ 2.60GHz with 16 cores on a x86\_64 hardware architecture with GNU/Linux operating system. The compiler is GNU Fortran (GCC) 4.8.5 20 (Red Hat 4.8.5-28) with the default optimization flags.    

AGP-CI calculations in Sec. \ref{sec:Energy}: performed in parallel on a workstation with Intel Xeon(R) CPU E3-1270 v6, with 8 cores, each at 3.80GHz on a x86\_64 hardware architecture with GNU/Linux operating system. PGI-15 compiler with the following flags: "-O4 -Mvect -Mprefetch -Mconcur=allcores -Mcache\textunderscore align -fast -fastsse". 

\bibliographystyle{apsrev4-1}
\bibliography{main}

\end{document}